\documentclass[12pt]{article}
\usepackage{html}
\usepackage{epsf}
\usepackage{graphicx}
\usepackage{amssymb}
\usepackage{hyperref}
\begin{document}
\title{MATTERS OF GRAVITY, The newsletter of the APS Topical Group on 
Gravitation}
\begin{center}
{ \Large {\bf MATTERS OF GRAVITY}}\\ 
\bigskip
\hrule
\medskip
{The newsletter of the Topical Group on Gravitation of the American Physical 
Society}\\
\medskip
{\bf Number 46 \hfill December 2015}
\end{center}
\begin{flushleft}
\tableofcontents
\vfill\eject
\section*{\noindent  Editor\hfill}
David Garfinkle\\
\smallskip
Department of Physics
Oakland University
Rochester, MI 48309\\
Phone: (248) 370-3411\\
Internet: 
\htmladdnormallink{\protect {\tt{garfinkl-at-oakland.edu}}}
{mailto:garfinkl@oakland.edu}\\
WWW: \htmladdnormallink
{\protect {\tt{http://www.oakland.edu/?id=10223\&sid=249\#garfinkle}}}
{http://www.oakland.edu/?id=10223&sid=249\#garfinkle}\\

\section*{\noindent  Associate Editor\hfill}
Greg Comer\\
\smallskip
Department of Physics and Center for Fluids at All Scales,\\
St. Louis University,
St. Louis, MO 63103\\
Phone: (314) 977-8432\\
Internet:
\htmladdnormallink{\protect {\tt{comergl-at-slu.edu}}}
{mailto:comergl@slu.edu}\\
WWW: \htmladdnormallink{\protect {\tt{http://www.slu.edu/colleges/AS/physics/profs/comer.html}}}
{http://www.slu.edu//colleges/AS/physics/profs/comer.html}\\
\bigskip
\hfill ISSN: 1527-3431

\bigskip

DISCLAIMER: The opinions expressed in the articles of this newsletter represent
the views of the authors and are not necessarily the views of APS.
The articles in this newsletter are not peer reviewed.

\begin{rawhtml}
<P>
<BR><HR><P>
\end{rawhtml}
\end{flushleft}
\pagebreak
\section*{Editorial}

Matters of Gravity has adopted a new publication schedule: it will appear in December and June.  The 
purpose of this change is so that a preliminary description of the GGR sessions of each upcoming April APS meeting can  be shown to the GGR membership before the deadline for submission of an abstract for the April meeting.
The next newsletter is due June 2016.  This and all subsequent
issues will be available on the web at
\htmladdnormallink 
{\protect {\tt {https://files.oakland.edu/users/garfinkl/web/mog/}}}
{https://files.oakland.edu/users/garfinkl/web/mog/} 
All issues before number {\bf 28} are available at
\htmladdnormallink {\protect {\tt {http://www.phys.lsu.edu/mog}}}
{http://www.phys.lsu.edu/mog}

Any ideas for topics
that should be covered by the newsletter should be emailed to me, or 
Greg Comer, or
the relevant correspondent.  Any comments/questions/complaints
about the newsletter should be emailed to me.

A hardcopy of the newsletter is distributed free of charge to the
members of the APS Topical Group on Gravitation upon request (the
default distribution form is via the web) to the secretary of the
Topical Group.  It is considered a lack of etiquette to ask me to mail
you hard copies of the newsletter unless you have exhausted all your
resources to get your copy otherwise.

\hfill David Garfinkle 

\bigbreak

\vspace{-0.8cm}
\parskip=0pt
\section*{Correspondents of Matters of Gravity}
\begin{itemize}
\setlength{\itemsep}{-5pt}
\setlength{\parsep}{0pt}
\item Daniel Holz: Relativistic Astrophysics,
\item Bei-Lok Hu: Quantum Cosmology and Related Topics
\item Veronika Hubeny: String Theory
\item Pedro Marronetti: News from NSF
\item Luis Lehner: Numerical Relativity
\item Jim Isenberg: Mathematical Relativity
\item Katherine Freese: Cosmology
\item Lee Smolin: Quantum Gravity
\item Cliff Will: Confrontation of Theory with Experiment
\item Peter Bender: Space Experiments
\item Jens Gundlach: Laboratory Experiments
\item Warren Johnson: Resonant Mass Gravitational Wave Detectors
\item David Shoemaker: LIGO Project
\item Stan Whitcomb: Gravitational Wave detection
\item Peter Saulson and Jorge Pullin: former editors, correspondents at large.
\end{itemize}
\section*{Topical Group in Gravitation (GGR) Authorities}
Chair: Deirdre Shoemaker; Chair-Elect: 
Laura Cadonati; Vice-Chair: Peter Shawhan. 
Secretary-Treasurer: Thomas Baumgarte; Past Chair:  Beverly Berger;
Members-at-large:
Andrea Lommen, Jocelyn Read, Steven Drasco, Sarah Gossan, Tiffany Summerscales, Duncan Brown, Michele Vallisneri.
Student Member: Jessica McIver.
\parskip=10pt

\vfill\eject

\section*{\centerline
{Remembering Jacob Bekenstein}}
\addtocontents{toc}{\protect\medskip}
\addtocontents{toc}{\bf GGR News:}
\addcontentsline{toc}{subsubsection}{
\it  Remembering Jacob Bekenstein, by Tsvi Piran}
\parskip=3pt
\begin{center}
Tsvi Piran, The Hebrew University of Jerusalem
\htmladdnormallink{tsvi.piran-at-mail.huji.ac.il}
{mailto:tsvi.piran@mail.huji.ac.il}
\end{center}

Rarely, very rarely a PhD thesis changes the way we view the world: Jacob Bekenstein's thesis was one of those. Beforehand black holes absorbed everything and emitted nothing. Afterwards, black holes attained entropy and temperature (the Bekenstein-Hawking temperature, with a nice appropriate acronym - BH) and the cold black holes became a bit warmer in many senses.  As a beginning graduate student, I was mesmerized, like many, by black holes. Fascinated by these new concepts and being in the then remote Israel, I asked Jacob for a copy of his thesis.  I remember my excitement when it arrived.  I was astonished to discover that a large fraction of it dealt with a proof of the no hair theorem. I wondered: ``isn't the discovery of the generalized law of thermodynamics worth a PhD on its own?''

Revolutionary ideas are never accepted easily.  Young Jacob, still a PhD student, had to struggle with many who disliked the new concept. I vividly recall a distinguished scientist who dismissed all this as ``a mere dimensional coincidence.'' By now we know that these ideas were the precursors of Hawking radiation and the very first signs of an essential link between general relativity and quantum physics.  The importance of the black hole's area that emerged from these ideas is yet another precursor of the holographic principle that occupies now a central role in different branches of physics. 

Then, in 1974, the young and Zionist Jacob made another pioneering move. Jacob who could have obtained a job in any top-ten University moved to the young Ben Gurion University, established just five years earlier in Be'er Sheva. Jacob built his family in Be'er Sheva and spent the next 15 years there before moving to the Hebrew University of Jerusalem in 1990. 

Jacob continued to produce remarkable and original work both in Be'er Sheva and in Jerusalem. Among those is the Bekenstein entropy bound that sets an upper bound on the amount of entropy that can be stored in a system that has a given amount of energy. The entropy bound led to yet another vivid controversy. The exchange took place in the form of a series of Physical Review papers by Jacob on one hand and Bill Unruh and Bob Wald on the other (all three having been fellow students at Princeton a decade earlier). The series of more and more elaborate gedanken experiments reminded me of the famous Einstein-Bohr correspondence. 

When MOND came out Jacob Bekenstein set out to offer (first with Mordechai Milgrom) a more elegant version of it.  Later on his own, he formulated TEVES, a relativistic theory that is a framework that accounts for observational effects that are associated with both dark matter and dark energy, without invoking either one of them. Jacob considered this last contribution as one of his most important ones. In an interview with a Haaretz journalist, he said: ``I think I'm still doing important things, but it will take more time for the public, and even important scientists, to understand what I did.'' 

Jacob's works and contributions are widely recognized. He was a member of the Israeli Academy and he has received the Israel prize, the Rothschild prize, the Weizmann prize, the Wolf prize and the Einstein prize of the APS.

One cannot write about Jacob without mentioning his religious faith, and the fact that he practiced it in a way that promoted his scientific work, rather than interfering with it. ``Particularly as an observant Jew, I'm interested in knowing how nature works'' he said to the Haaretz interviewer. 
He was also a strict vegetarian, adopting this stance on moral grounds long before it became a fad.  Jacob was extremely modest and in spite of his international stature he never had any demands or requests. His office furniture was the simplest on our floor. 

Jocob Bekenstein passed away in August, 2015 while visiting the University of Helsinki in Finland.  He is survived by his wife, Bilha, three children, all of them scientists, and seven grandchildren. He will be greatly missed by his family, his colleagues at the Hebrew University and worldwide.

\section*{\centerline
{we hear that \dots}}
\addtocontents{toc}{\protect\medskip}
\addcontentsline{toc}{subsubsection}{
\it we hear that \dots , by David Garfinkle}
\parskip=3pt
\begin{center}
David Garfinkle, Oakland University
\htmladdnormallink{garfinkl-at-oakland.edu}
{mailto:garfinkl@oakland.edu}
\end{center}

Henriette Elvang has received the APS Maria Goeppert Mayer Award

Vicky Kalogera has received the APS Hans Bethe Prize

Pablo Laguna has received the APS Bouchet Award

David Tanner has received the APS Frank Isakson Prize

Emanuele Berti, Laura Cadonati, Yanbei Chen, Dennis Coyne, Daniel Sigg, Mark Trodden, Alan Weinstein, and Bernard Whiting have been elected APS Fellows.

Hearty Congratulations!

\section*{\centerline
{100 years ago}}
\addtocontents{toc}{\protect\medskip}
\addcontentsline{toc}{subsubsection}{
\it 100 years ago, by David Garfinkle}
\parskip=3pt
\begin{center}
David Garfinkle, Oakland University
\htmladdnormallink{garfinkl-at-oakland.edu}
{mailto:garfinkl@oakland.edu}
\end{center}

In November, 1915 Einstein (finally!) came up with General Relativity in its current form, including the correct form of the field equations.  He presented the new theory in a lecture on Nov. 25, and subsequently wrote it up in a paper.  For the Nov. 15 lecture (in English translation) see

\htmladdnormallink 
{\protect {\tt {http://einsteinpapers.press.princeton.edu/vol6-trans/129?ajax}}}
{http://einsteinpapers.press.princeton.edu/vol6-trans/129?ajax}

while for the subsequent paper (also in English translation) see

\htmladdnormallink 
{\protect {\tt {http://einsteinpapers.press.princeton.edu/vol6-trans/158?ajax}}}
{http://einsteinpapers.press.princeton.edu/vol6-trans/158?ajax}

Also in 1915, Hilbert came up with the action for general relativity. 

\vfill\eject
\section*{\centerline
{GGR program at the APS April meeting in Salt Lake City}}
\addtocontents{toc}{\protect\medskip}
\addcontentsline{toc}{subsubsection}{
\it GGR program at the APS meeting in Salt Lake City , by David Garfinkle}
\parskip=3pt
\begin{center}
David Garfinkle, Oakland University
\htmladdnormallink{garfinkl-at-oakland.edu}
{mailto:garfinkl@oakland.edu}
\end{center}

We have a very exciting GGR related program at the upcoming APS April meeting at Salt Lake City, UT.  Our Chair-elect, Laura Cadonati did an excellent job of putting together this program.
\vskip0.25truein

{\bf Note that the deadline for submitting an abstract for this meeting is Friday, January 8, 2016 at 5:00 pm EST}

abstracts can be submitted at

\htmladdnormallink 
{\protect {\tt {http://abstracts.aps.org}}}
{http://abstracts.aps.org}

The invited sessions sponsored by GGR are as follows:\\

{\bf First Results from Advanced LIGO}\\
Peter Fritschel, The Advanced LIGO Detector \\
Chad Hanna, Searches for GW transients in Advanced LIGO \\
Salvatore Vitale, First Results from Advanced LIGO: Parameter estimation and tests of GR \\

{\bf The Future of Ground Based Gravitational Wave Detection}\\
Alessandra Corsi, The future of ground based GW astrophysics \\
Stefan Ballmer, The Emerging Gravitational-wave Detection Network \\
Sheila Dwyer, Future Detector Development  \\

{\bf Schwarzschild Black Hole}\\
Marcia Bartusiak, Schwarzschild solution: an historical prospective \\
Aycin Ay, Black Hole observations \\
Javier Olmedo, Quantum black holes in Loop Quantum Gravity\\

{\bf Measuring Big G}\\
Stephan Schlamminger, Recent measurements of G \\
(speaker TBD), Atomic Interferometry, summary and results \\
Christian Rothleitner, New routes to measuring Big G \\

{\bf Pulsar Timing Array}\\
(joint with DAP)\\
Maura McLaughlin, Building A Galactic Scale Gravitational Wave Observatory \\
Xavier Siemens, primordial GWs \\
Sarah Burke Spolaor, Measuring Evolution in the Universe with Pulsar Timing Arrays \\

{\bf Space Based Gravitational Wave Astrophysics}\\
(joint with DAP)\\
Martin Hewitson, LISA Pathfinder: picometers and femtoNewtons in space \\
Scott Hughes, Listening to the low-frequency gravitational-wave band \\
(speaker TBD), mining GW and EM data to maximize science return \\

{\bf Theory Developments in Particle Physics and Gravitation}\\
(joint with DPF)\\
John Preskill, Quantum Information and its Connections with Black Holes \\
Andrew Strominger, Heineman Prize Talk: Tying Soft Theorems to Symmetries in Gravity and Gauge Theory \\ 
Veronica Hubeny, Gravitational lessons from Holographic Entanglement Entropy \\ 

{\bf Computing Meets Experiment in Gravity}\\
(joint with DCOMP)\\
Ian Harry, Computing in large scale experiments \\
Patricia Schmidt, Computing in gravity theory \\
Nepomuke Otte, High performance computing in astroparticle physics: very-high-energy gamma-ray observations with imaging atmospheric Cherenkov telescopes \\

\vfill\eject
\section*{\centerline
{Hawking radiation}}
\addtocontents{toc}{\protect\medskip}
\addtocontents{toc}{\bf Conference reports:}
\addtocontents{toc}{\protect\medskip}
\addcontentsline{toc}{subsubsection}{
\it Hawking radiation, 
by Larry Ford}
\parskip=3pt
\begin{center}
Larry Ford, Tufts University 
\htmladdnormallink{ford-at-cosmos.phy.tufts.edu}
{mailto:ford@cosmos.phy.tufts.edu}
\end{center}

On Aug. 24-29, 2015 there was a Nordita conference on Hawking radiation, which took place 
at the KTH Royal Institute of Technology, 
in Stockholm Sweden. It was organized by L. Mersini-Houghton, Y-C Ong, and M. Perry, 
and supported by Nordita and the University of North Carolina. The program
featured 17 talks, which allowed ample time for discussion, and some informal discussion 
sessions. A central theme was the black hole information problem, but many other aspects 
of the Hawking effect and related topics were also addressed.  

One class of proposals would eliminate the black hole information problem by effectively 
eliminating black holes altogether. Two such ideas were presented by G. 't Hooft and by 
L. Mersini-Houghton. 't Hooft presented a model in which gravity may be viewed as a theory 
with spontaneously broken, local conformal symmetry. In this model, a black hole seems 
to become a regular soliton without a singularity, horizon, or firewalls. Mersini-Houghton 
described a model of backreaction in which an ingoing negative energy flux is completely 
absorbed by the matter in a collapsing star. In this model, the backreaction is sufficiently 
strong to prevent the formation of either the horizon or a singularity. As a counter point to the 
latter suggestion, J. Bardeen reviewed the more conventional and conservative viewpoint of 
backreaction as described by the renormalized energy momentum tensor of the quantum field 
on the background geometry of the collapse spacetime. In this viewpoint, the backreaction is 
small for large black holes and does not prevent the formation of the horizon. Rather, it describes 
the gradual loss of mass by the black hole as it evaporates.

E. Mottola discussed a model of ``gravitational condensate stars'' using Schwarzscild's interior 
solution with negative pressure. In this model, the region within the black hole horizon is replaced 
by deSitter spacetime and is non-singular. C. Rovelli and F. Vidotto gave two related talks on the 
possibility of black hole to white hole tunneling. They discussed a classical spacetime in which matter 
falls into a black hole and then emerges from a white hole. They argued that there is a nonzero probability 
in quantum gravity for a black hole to tunnel into a white hole, and the tunneling rate was estimated. 
Possible observable effects of this process were discussed.

The talks of S. Hawking and M. Perry presented a proposed solution to the information problem, 
the essence of which is that information which falls into a black hole is not really lost, but rather encoded 
in perturbations of the horizon. This proposal is a modification of a description of information radiated in 
gravity waves, developed by Bondi, Metzner, and Sachs (BMS). The BMS approach describes the 
perturbations of future null infinity due to outgoing gravity waves in terms of a group of supertranslations, 
the BMS group. The present proposal involves an analogous description of the perturbations of the future 
horizon of a black hole spacetime by infalling matter or gravity waves. This leads to a type of ``black hole memory'' 
in which a formal record remains of everything which ever crossed the horizon. The key unanswered question 
is whether this formal record can later be accessed in a way that solves the information problem. Skepticism on 
this point was expressed by some participants, based on the exponential decay of black hole perturbations as 
seen by outside observers.

The firewall solution to the black hole information problem, in which an infalling observer is destroyed 
at the horizon was mentioned in several talks. It was most directly addressed by J. Louko, who presented 
a model of a black hole firewall as the Rindler vacuum in two dimensional flat spacetime, where the 
energy density is singular on the Rindler horizon. It was found that the response of an Unruh-DeWitt 
detector is finite as it crosses this horizon, and that the entanglement of a pair of such detectors can 
survive passage through this model firewall. At least in this simplified model, the firewall is not strong 
enough to destroy the infalling observer, or to provide a solution to the information problem.
 
There were two presentations which dealt with decoherence and related issues. C. Kiefer treated 
black holes as open quantum systems. Two models were discussed. One model involved a set of 
quantum harmonic oscillators. One of these oscillators models the Hawking radiation, and the other 
the black hole, and the entanglement between these was studied. The other model is a two 
dimensional model in which Hawking radiation leads to decoherence of black hole superpositions. 
L. Stodolsky also discussed decoherence and its possible role in black hole physics, drawing on 
examples of other types of quantum systems.

Other aspects of black hole physics were addressed in the talks of B. Whiting, F. Dowker, and 
K. Stelle. Whiting compared and contrasted classical and quantum calculations on black hole 
spacetimes. He noted that the classical problem of the solution of wave equations and calculation 
of self force uses many of the same techniques which can be used to compute the backreaction 
of the Hawking radiation on a black hole. He also summarized some of the technical and conceptual 
problems which are unique to the quantum case. Dowker discussed the generalized second law of 
thermodynamics, including the inputs to its proof and the broader implications of the generalized 
second law of thermodynamics for the unity of physics. Stelle discussed black hole solutions 
in higher derivative gravity theory.
 
The talks of L. Parker, C. Misner, and M. Duff dealt with some topics which are broader than black 
hole physics, but  nonetheless potentially relevant to Hawking radiation. Parker gave a review of 
cosmological particle creation. This effect, first treated treated by Parker, laid the groundwork for 
Hawking's 1975 paper on particle creation by black holes. Misner discussed some physical 
examples of quantum gravity effects, including the quantum limits on gravitational radiation 
at very low rates. One example is a human being exercising. Classically, a very low flux of gravity 
waves is produced. However, the chances of emitting even a single graviton during a work out 
is very low, so this is a case where quantum gravity in principle has an effect on a macroscopic 
scale. The talk of M. Duff described how two copies of a Yang-Mills theory can reproduce the 
symmetries and linearized equations of gravity theory, leading to a formalism in which the 
gravitational field is viewed as the square of Yang-Mills theory.

On the last day of the conference, S. Fulling led a discussion on several possibilities for 
resolving the information puzzle. Some of the options include: formation of remnants, a 
final burst, firewalls, fuzzballs, formation of baby universes into which information flows, 
fireworks (meaning a dramatic effect which prevents black hole formation), and finally 
the view that information is lost and black hole evaporation is not unitary. No clear consensus 
arose in the discussion, despite Fulling's efforts to find agreement on at least some issues. 
This indicates that many aspects of Hawking radiation remain controversial. 

Conference summaries, or viewpoints, were presented by P. Davies and by S. Hawking.  
\vfill\eject

\section*{\centerline
{Quantum Information in Quantum Gravity II}}
\addtocontents{toc}{\protect\medskip}
\addcontentsline{toc}{subsubsection}{
\it Quantum Information in Quantum Gravity II, 
by Rob Myers}
\parskip=3pt
\begin{center}
Rob Myers, Perimeter Institute
\htmladdnormallink{rmyers-at-perimeterinstitute.ca}
{mailto:rmyers@perimeterinstitute.ca}
\end{center}

In recent years, it has become widely appreciated that quantum information theory is a fruitful lens with which to examine the conundrums of quantum gravity. As the conference title indicates, this meeting was the second in a series of workshops aimed at examining recent developments on this front and facilitating interactions between the quantum information and the string/quantum gravity communities, as well as researchers with common interests studying condensed matter physics and quantum field theory.  
The first of these meetings was held in August 2014 at the University of British Columbia. I was so enthused by all of the talks and my discussions there that the idea of holding a second meeting was born on the plane flight home from Vancouver. Hence I recruited Mark Van Raamsdonk at UBC and Guifre Vidal at Perimeter to help me organize the second workshop and it brought together approximately 70 enthusiastic participants for a stimulating week of seminars and discussions at the Perimeter Institute, August 17-21, 2015.

The roots of the interplay between quantum information and quantum gravity stretch back to 1973, a time when rapid progress was being made in uncovering the mysteries of black holes but quantum information theory was yet to be born as a field of research. In the spring of that year, Jacob Bekenstein, then a 25 year-old graduate student finishing up his PhD at Princeton, came up with the radical proposal that black holes should have an intrinsic entropy and that this entropy was given by the area of the event horizon measured in units of the Planck scale. His surprising suggestion hinted that rather than just being elegant and pristine solutions of general relativity, physical black holes were actually endowed with an enormous but unseen microscopic complexity. While his idea originally met with strong opposition, this melted away two years later with Stephen Hawking's discovery that black holes emit (nearly) thermal radiation. The Bekenstein-Hawking formula, S=A/4G, is now widely regarded as one of the most remarkable discoveries in fundamental physics.  
Now with our intuition from thermodynamics, entropy is typically thought of as a measure of disorder. However, precisely the same mathematical quantity also plays a central role in information theory; for example, entropy measures the average number of bits needed to store one ``letter'' of a message. Recently, Bekenstein and Hawking's discoveries have been recognized as the first clues of profound connections between gravity and information theory, in particular its modern incarnation as quantum information theory. 

A natural framework where much of this interplay between gravity and information has been developing in recent years has been the AdS/CFT correspondence. Recall that this correspondence sets up an equivalence between gravity in anti-de Sitter space (AdS) and a special class of quantum field theories, called conformal field theory (CFT), in one less dimension. Since this correspondence relates theories in different dimensions, it is often referred to as “holography.” A topic currently being intensively studied is holographic entanglement entropy. In general, entanglement entropy provides a measure of the subtle correlations between the microscopic degrees of freedom in a quantum theory. It is a concept that was broadly applied in quantum information, where these entanglements are a resource that can be used for ultrafast computing or ultrasecure communications, but it also plays a role in condensed matter physics, as well as quantum field theory and quantum gravity. According to a prescription originally proposed jointly by Shinsei Ryu, a condensed matter theorist, and Tadashi Takayanagi, a string theorist, holography encodes entanglement entropy of the boundary theory in the geometry of the bulk spacetime of the dual gravity theory. In fact, one evaluates the Bekenstein-Hawking formula on extremal surfaces in the bulk which are homologous to the corresponding boundary region. While we are still trying to understand the full implications of this remarkable result, these ideas motivated Mark Van Raamsdonk to propose that entanglement of the microscopic degrees of freedom should be a key feature in the emergence of spacetime in quantum gravity. More recently, Mark and his collaborators have shown that in a holographic setting, basic constraints coming from the properties of entanglement entropy in the boundary theory can be translated into the requirement that the bulk spacetime geometry must satisfy the Einstein equations, at least at the linearized level. Hence we are beginning to see that entanglement seems to play a role in realizing Einstein’s view of  gravity, in which  spacetime is not only a stage on which physical phenomena take place but also a dynamical player as the agent of the gravitational force. 

The discussion above only hints at a few of the fascinating ideas which have emerged in recent years and which set the stage for our August meeting. With twenty-five talks, as well as a number of short ``gong show'' presentations and posters, I can't give a proper description of all of the new work that was discussed at the meeting and so let me begin with a couple of highlights from my personal perspective:  Ted Jacobson gave a wonderful description of his new argument that derives Einstein's equations from some simple assumptions about entanglement entropy in small regions of spacetime. While these assumptions are still being scrutinized, Ted's construction points to a profound connection between entanglement and Einstein gravity in a general setting: no holography needed here! 
Matt Headrick discussed an interesting new perspective of holographic entanglement entropy, which he is developing with Michael Freedman. Their approach describes the entanglement in terms of ``bit threads'' which one might see as mapping out a flow of correlations through the bulk spacetime. These threads have a thickness of the order of the Planck scale and so the extremal surfaces of Ryu and Takayanagi are bottlenecks which limit the flows established by the threads. Matt's talk had me thinking back to Maxwell's mechanical model of the electromagnetic field and so perhaps his ``bit lines'' are hints of a new entanglement formulation of Einstein's equations.

One interesting connection between geometry and entanglement has been recognized in the description of discrete systems, e.g., of spins. In particular, the ground state wave-function of many such systems can be accurately and efficiently described by tensor networks, which can be regarded as a form of the quantum circuits that are prevalent in quantum information discussions. The Multiscale Entanglement Renormalization Ansatz (MERA) is a class of such tensor networks that was originally developed by Guifre Vidal and gives a useful representation of critical systems (i.e., systems that flow to a CFT in the infrared).  In 2009, Brian Swingle conjectured that MERA gives a tensor network realization of the AdS/CFT correspondence. Hence this meeting gave considerable attention to tensor networks and their possible connections to holography. Glen Evenbly described his recent work with Guifre on understanding renormalization group transformations in lattice models and how this approach naturally converts a discrete path integral representation of the ground state into a MERA network. Jutho Haegemann reviewed his past work on cMERA, a MERA construction for continuum quantum field theories. James Sully and Bartek Czech both gave energetic talks describing their recent work which connects the AdS/CFT correspondence with MERA networks through an integral transform. In their construction, the tensor network is more directly related to a de Sitter (rather than anti-de Sitter) geometry. Intriguingly, as described by Michal Heller, a similar de Sitter geometry seems to arise in the description of entanglement in any conformal field theory in any number of dimensions.

Another theme at the workshop was the exploration of additional quantum information tools and how they might arise in holography. Tensor networks provided an exciting example of these explorations in the talk by Fernando Pastawski. Fernando, Benni Yoshida and John Preskill, all quantum information theorists, recently joined forces with Dan Harlow, a string theorist, to build some intriguing toy models of holography using ideas from quantum error correction. Their models exhibit many of the interesting features of the AdS/CFT correspondence. For example, the Ryu-Takayanagi formula is exactly reproduced for any connected boundary region, with a minimal cut through the tensor network. Xiao-Liang Qi (a condensed matter theorist) described how he has been extending these models with his collaborators. Tadashi Takayanagi and Nima Lashkari both described their independent investigations into the question of what the holographic description of a quantity known as the Fisher information metric might be. Further, Adam Brown gave a fascinating talk on recent work that he and his collaborators in Stanford are developing on how the complexity of the boundary state might be measured by evaluating the action of a particular region of the bulk spacetime, which he dubbed the “Wheeler-deWitt” patch. 

More information on the conference, as well as full recordings of all of the talks, can be found at: 

\htmladdnormallink 
{\protect {\tt {https://www.perimeterinstitute.ca/conferences/quantum-information-quantum-gravity-ii}}}
{https://www.perimeterinstitute.ca/conferences/quantum-information-quantum-gravity-ii}

As well as the exciting new ideas, the energetic and enthusiastic atmosphere at the meeting came from the good number of students and postdocs who attended. We coordinated the dates of our meeting with those of the Mathematica Summer School on Entanglement, which was hosted at Perimeter Institute the following week (August 24-29, 2015). Many of these younger researchers were able to stay on to attend both meetings. Some of the young-at-heart participants, namely Horacio Casini, Guifre Vidal and myself, also stayed for the following week to serve as lecturers at the school.

We were devastated to learn, on the first day of the meeting, that Jacob Bekenstein had just passed away. It was  profoundly sad to hear the news that we had lost a true pioneer.  However, we were somewhat consoled by the fact that our meeting was itself a reminder of how the seeds he planted so long ago have taken hold and flourished in the intervening years. We dedicated the meeting to him, and hopefully, planted new seeds that will bear fruit in the years to come. 
\vfill\eject

\section*{\centerline
{Gravity-new perspectives from strings and higher dimensions}}
\addtocontents{toc}{\protect\medskip}
\addcontentsline{toc}{subsubsection}{
\it Gravity-new perspectives from strings and higher dimensions, 
by Simon Ross}
\parskip=3pt
\begin{center}
Simon Ross, University of Durham
\htmladdnormallink{S.F.Ross-at-durham.ac.uk}
{mailto:S.F.Ross@durham.ac.uk}
\end{center}

A workshop on “Gravity - new perspectives from strings and higher dimensions” was held in the lovely setting of the Centro de Ciencias de Benasque Perdo Pascual, in the Spanish Pyrenees, from the 12th to the 24th of July 2015. This was the fourth in a continuing series of workshops on gravity at the centre, and continued a strong tradition of combining interesting talks with lively discussion. The programme included a large number of talks by younger researchers, underlining the continuing vitality of the field.  Slides from the talks are available at

\htmladdnormallink 
{\protect {\tt {http://www.benasque.org/2015gravity/cgi-bin/talks/allprint.pl}}}
{http://www.benasque.org/2015gravity/cgi-bin/talks/allprint.pl}

There was a lot of discussion of holography and AdS/CFT. The meeting opened with a talk by David Mateos on applications of holography to QCD, describing progress on attempts to construct a colour superconductor holographically. Later Wilke van der Schee discussed the modelling of nuclear collisions by colliding shocks in AdS, explaining how a model with a non boost-invariant rapidity profile could better fit observations. There were two talks describing different approaches to effective actions for hydrodynamics, by Felix Haehl and Jan de Boer. The first explained how all the non-dissipative classes of transport can be captured by an action based on a Schwinger-Keldysh doubling of the degrees of freedom, while the latter discussed a holographic construction of the action from a double Dirichlet problem in the bulk. Nikolay Bobev discussed the calculation of the free energy of supersymmetric field theories on the sphere from a bulk perspective, reproducing results obtained from localisation in the field theory. Javier Mas used collapsing matter in AdS as a model to study the appearance of collapses and revivals (periodic behaviour in the expectation values of field theory observables).

There was a particular focus on the role of entanglement entropy in holography, with a large number of talks on aspects of this question. The first, by Michal Heller, described how in a 1+1 field theory a partial order can be put on the space of spatial intervals, giving it the causal structure of de Sitter space, and showed that the change in the entanglement entropy associated to each interval produced by a perturbation obeys the scalar wave equation in this space. Alexandre Belin discussed the calculation of Renyi entropies, which are useful in calculating entanglement entropy via the ``replica trick'', and showed that there is a limitation to this by exhibiting a case where the Renyi entropies are not an analytic function of the number of replicas. Chris Herzog discussed two calculations showing the importance of boundary contributions in the calculation of entanglement entropy. Pablo Bueno discussed the case where the entangling surface has a sharp corner, and argued that this has a universal structure in the limit where the corner angle is nearly $\pi$. Aaron Wall discussed how the generalized second law and the entanglement entropy motivate a generalised notion of entropy, and used it to formulate a quantum version of the focussing conjecture. Aitor Lewkowycz discussed work in progress generalizing his proof of the Ryu-Takayanagi conjecture with Maldacena to construct a proof of the covariant HRT conjecture. Henry Maxfield argued that three dimensional bulks provide a particularly useful laboratory for studying entanglement holographically. There were two talks on seeing holographic features in more general quantum systems: Max Rota discussed various measures of entanglement in qubit systems, and the extent to which generic systems have the monogamy property seen holographically. Will Kelly showed that a suppression of corrections to the entanglement entropy from excitations supported on one side of the entangling surface seen holographically can be reproduced directly in the field theory. Finally, there were two entirely non-holographic talks by Guifre Vidal reviewing the multiscale entanglement renormalization ansatz (MERA) and its use in lattice systems. These provided a very useful introduction to a technology with intruiging similarities to the holographic calculations.

There was also an interesting selection of talks on other aspects of gravity and black holes. Pau Figueras gave a general review of work on stability of black holes, and discussed recent numerical work on the (in)stability of higher-dimensional black holes and black rings. Benson Way discussed stability in AdS, and described some new black hole solutions in AdS with resonant wave excitations outside the horizon. Roberto Emparan led a lively discussion of the large $D$ limit in general relativity, explaining how the description of black holes simplifies in this limit. Finn Larsen discussed logarithmic corrections to the black hole entropy in supergravity theories, arguing that unlike for non-supersymmetric theories, the coefficient of the logarithm is universal and independent of the specific black hole considered. Emil Martinec closed the meeting with an interesting discussion of black hole microstates, arguing that horizon formation could be related to condensation of hypermultiplets in a dual quiver gauge theory, and that in near-extremal solutions most of the black hole degrees of freedom would be supported at the inner horizon.

There were also a few talks less directly connected to gravity. Andrea Puhm gave a talk early in the meeting on an effective action approach to brane - antibrane systems, and argued that from this point of view the addition of an antibrane does not introduce any pathologies. This was followed up by an informal discussion led by Iosif Bena on antibranes. Marcello Ortaggio talked about the classification of solutions to the Maxwell equations, and presented new results on solutions with vanishing scalar invariants.

The programme of invited talks was of exceptionally high quality, and left ample time for discussion and enjoying the delights of Benasque, from long lunches in the sunshine to hiking in the surrounding mountains. These meetings have become an established feature on the calendar for people in this area, and I look forward to the next one in 2017.

\end{document}